\documentclass[two column,showpacs,preprintnumbers,amsmath,amssymb]{revtex4}
\usepackage{graphicx}
\usepackage{dcolumn}
\usepackage{bm}

\begin{document}
\title{Optimal path of diffusion over the saddle point and fusion of massive nuclei}
\author{Chun-Yang Wang,$^1$ Ying Jia,$^2$ and Jing-Dong Bao$^{1,3}$}
\thanks{Corresponding author. Electronic mail: jdbao@bnu.edu.cn}
\affiliation{$^1$Department of Physics, Beijing Normal University,
Beijing 100875, China\\
 $^2$College of Science, The Central University for Nationalities, Beijing 100081, China\\
 $^3$Center of Theoretical Nuclear Physics, National Laboratory of Heavy-ion Accelerator, Lanzhou 730000, China}

\date{\today}

\begin{abstract}
Diffusion of a particle passing over the saddle point of a
two-dimensional  quadratic  potential is studied via a set of
coupled Langevin equations and the expression for the passing
probability is obtained exactly. The passing probability is found to
be strongly influenced by the off-diagonal components of inertia and
friction tensors. If the system undergoes
 the optimal path to pass over the saddle
point by taking an appropriate direction of initial velocity into
account, which departs from the potential valley and has minimum
dissipation, the passing probability should be enhanced. Application
to fusion of massive nuclei, we show that there exists the optimal
injecting choice for the deformable target and projectile nuclei,
namely, the intermediate deformation between spherical and extremely
deformed ones which enables the fusion probability to reach its
maximum.

\end{abstract}
 \maketitle

\section{\label{sec:level1}INTRODUCTION}

The saddle point passage problem  is of great interest in various
fields of physics, such as collision of molecular systems, atomic
clusters, biomolecules, and so on. The previous studies on this
issue were mostly concentrated on a simple diffusive dynamics with
single degree of freedom, where the Langevin equation  with constant
coefficients can be easily solved in the case of a quadratic
potential \cite{hof}. However, since many processes obviously
involve more than one degree of freedom, which the one-dimensional
(1D) model does not distinctly hold,  high  dimension at least two
dimension (2D) is necessary. A case in point would be the fusion
reaction of massive nuclei, where the fusion is induced by diffusion
\cite{ari,swi} and the asymmetrical or the neck degree of freedom of
compound nuclei needs to be considered \cite{agu1,agu2}. For these
systems with the contact point of two colliding nuclei being very
close to the conditional saddle point, the potential energy surface
(PES) around the saddle point can be approximated to be a quadratic
type. Under this approximation,
 Abe \emph{et al.} \cite{abe1} obtained an analytical
expression for the multi-dimensional saddle-point passing
probability. Some authors discussed quantum effect of the fusion
probability by using the real time path integral \cite{baonpa} or
the quantum transport equation \cite{tak1,tak2}, respectively.
Boilley \emph{et al.} \cite{boi} studied the influence of initial
distribution upon the passing probability. Anomalous diffusion
passing over the saddle point of 1D quadratic potential was also
discussed in Ref. \cite{bao1}. Nevertheless, it is not  completely
clear for the dynamical role of non-transport degrees of freedom.
This might be very important for the quasi-fission mechanism in the
fusion reaction, because the average path of the fusing system in a
multi-dimensional PES should be controlled by the off-diagonal
components of inertia and friction tensors before the system arrives
firstly at the conditional saddle point.

Recently, theoretical calculations for the fusion barrier
distribution, accounting for the surface curvature correction to the
nuclear potential, have been  presented by Hinde \emph{et al}.
\cite{hinde1,hinde2,hinde3}. The geometrical effect significantly
changes the near-barrier fusion cross section and the shape of the
barrier distribution through an angle-dependent potential, where the
target nucleus bears quadrupole and bexadecapole deformations while
the projectile one is in a spherical shape. In these calculations,
the surface curvature correction to the sphere-to-sphere nuclear
potential exerts influences upon the fusion probability through the
height of fusion barrier. However, the dynamical coupling effect of
various deformative degrees of freedom needs to be added from the
viewpoint of fusion by diffusion \cite{swi}.

The primary purpose of this paper is to study the influence of
coupling between two degrees of freedom upon the passing
probability. In Sec. II, we report the analytical expression of the
saddle-point passing probability by solving the 2D coupled Langevin
equation with constant coefficients. In Sec. III, we discuss the
effects of off-diagonal components of inertia, friction and
potential-curvature tensors and then determine the optimal diffusive
path. Sec. IV gives an application of this study to the actual
fusion process of massive nuclei. A summary is written in Sec. V.

\section{The passing probability}

We consider the directional diffusion of a particle in a 2D
quadratic PES: $U(x_1, x_2)=\frac{1}{2}\omega_{ij}x_ix_j$ with
$i,j=1,2$ and $\det\omega_{ij}<0$, the
 motion of the particle is described by the Langevin
equation
\begin{equation}
m_{ij}\ddot{x}_{j}(t)+\beta_{ij}\dot{x}_{j}(t)+\omega_{ij}x_{j}(t)=\xi_{i}(t)
\end{equation}
with $x_{j}(0)=x_{j0}$ and $\dot{x}_j(0)=v_{j0}$, where $x_{10}<0$
and $v_{10}>0$. Here and below the Einstein summation convention is
used. The two components of the random force are assumed to be
Gaussian white noises and their correlations obey the
fluctuation-dissipation theorem $
\langle\xi_{i}(t)\xi_{j}(t')\rangle=k_{_{B}}Tm_{ik}^{-1}\beta_{kj}\delta(t-t')$,
where $k_B$ is the Blotzmann constant and $T$ the temperature.

Assuming that $x_{1}$-axis is the transport  direction
[$\omega_{11}<0$], we write
  the reduced distribution
function of the particle for $x_1$ while the  variables $x_{2}(t)$,
$v_{1}(t)$ and $v_{2}(t)$ are integrated out,
\begin{eqnarray}
W(x_{1},t; x_{10},x_{20},v_{10},v_{20})
&=&\frac{1}{\sqrt{2\pi}\sigma_{x_{1}}(t)}\nonumber\\
&&\textrm{exp}\left({-\frac{(x_{1}(t)-\langle
x_{1}(t)\rangle)^{2}}{2\sigma^{2}_{x_{1}}(t)}}\right).\nonumber\\
\end{eqnarray}
Integrating over $x_{1}$ from zero to infinity, we determine the
passing probability over the saddle point [$x_1=x_2=0$] as
\begin{eqnarray}
P(t; x_{10},x_{20},v_{10},v_{20})&=&\int^{\infty}_{0}W(x_{1},t;
x_{10},x_{20},v_{10},v_{20})dx_{1}\nonumber\\
&=&\frac{1}{2}\textrm{erfc}\left(-\frac{\langle
x_{1}(t)\rangle}{\sqrt{2}\sigma_{x_{1}}(t)}\right).
\end{eqnarray}

Applying the Laplace transform technique to Eq. (1), we thus get
$x_{1}(t)$ and its variance $\sigma^{2}_{x_{1}}(t)$ at any time,
\begin{eqnarray}
x_{1}(t)&=&\langle
x_{1}(t)\rangle+\sum^{2}_{i=1}\int^{t}_{0}H_{i}(t-t')\xi_{i}(t')dt',\\
\sigma^{2}_{x_{1}}(t)&=&\int^{t}_{0}dt_{1}H_{i}(t-t_{1})\int^{t_{1}}_{0}dt_{2}\langle
\xi_{i}(t_{1})\xi_{j}(t_{2})\rangle H_{j}(t-t_{2}),\nonumber\\
\end{eqnarray}
where the mean position of the particle along the transport
direction is given by
\begin{equation}
\langle
x_{1}(t)\rangle=\sum^{2}_{i=1}\left[C_{i}(t)x_{i0}+C_{i+2}(t)v_{i0}\right],
\end{equation}
which
 relates to the initial position and velocity. The time-dependent factors in Eq. (6) with
 exponential forms according to the residual theorem  are $C_{i}(t)=\mathcal{L}^{-1}[\textsl{F}_{i}(s)/\textsl{P}(s)]$
$(i=1\ldots4)$, the two response functions in Eqs. (4) and (5) read
$\textsl{H}_{1}(t)=\mathcal{L}^{-1}[\textsl{F}_{5}(s)/\textsl{P}(s)]$
and
$\textsl{H}_{2}(t)=\mathcal{L}^{-1}[\textsl{F}_{6}(s)/\textsl{P}(s)]$,
where $\mathcal{L}^{-1}$ denotes the inverse Laplace transform. The
expressions of $P(s)$ and $F_{i}(s)$ $(i=1,...,6)$  are written in
the appendix.

\section{The optimal diffusive path}

\subsection{The coupling effect of two degrees of freedom}

\begin{figure}
\includegraphics[scale=0.4]{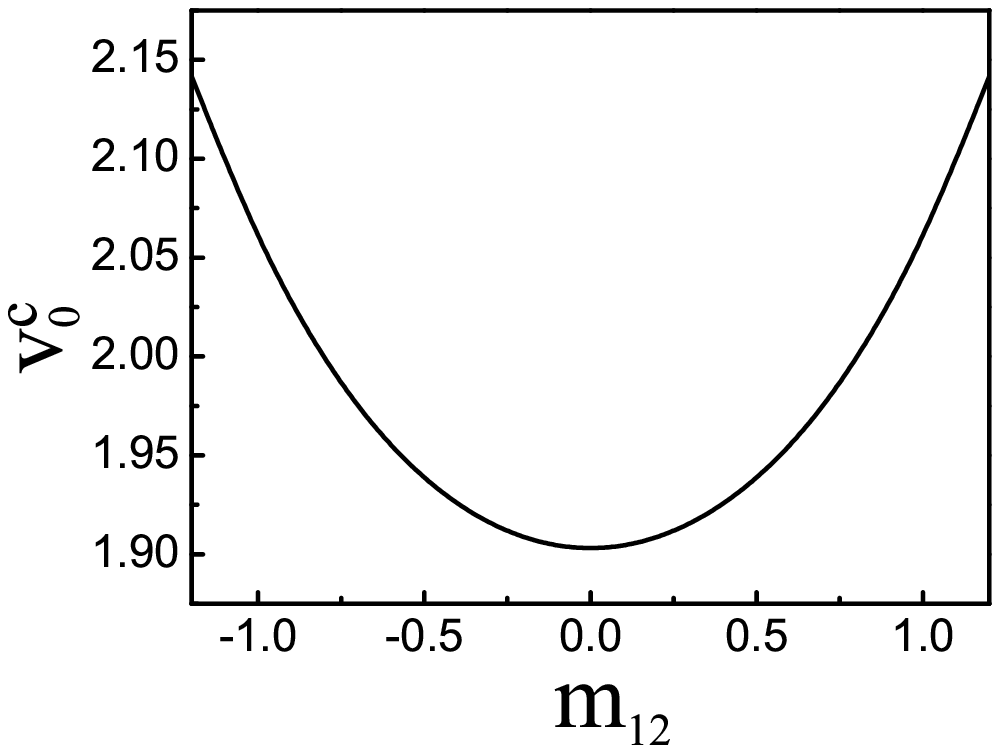}\includegraphics[scale=0.42]{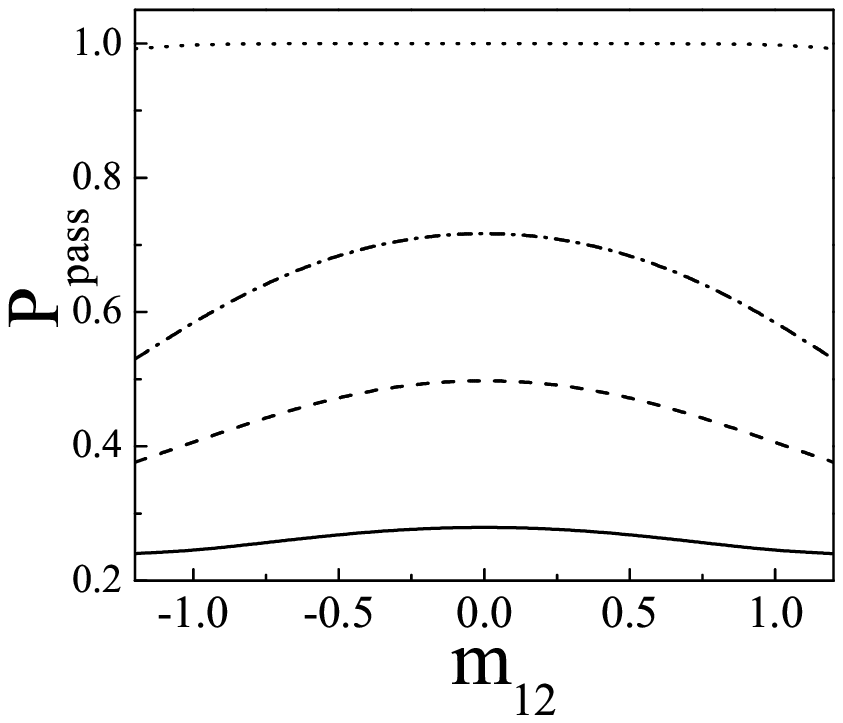}
\includegraphics[scale=0.4]{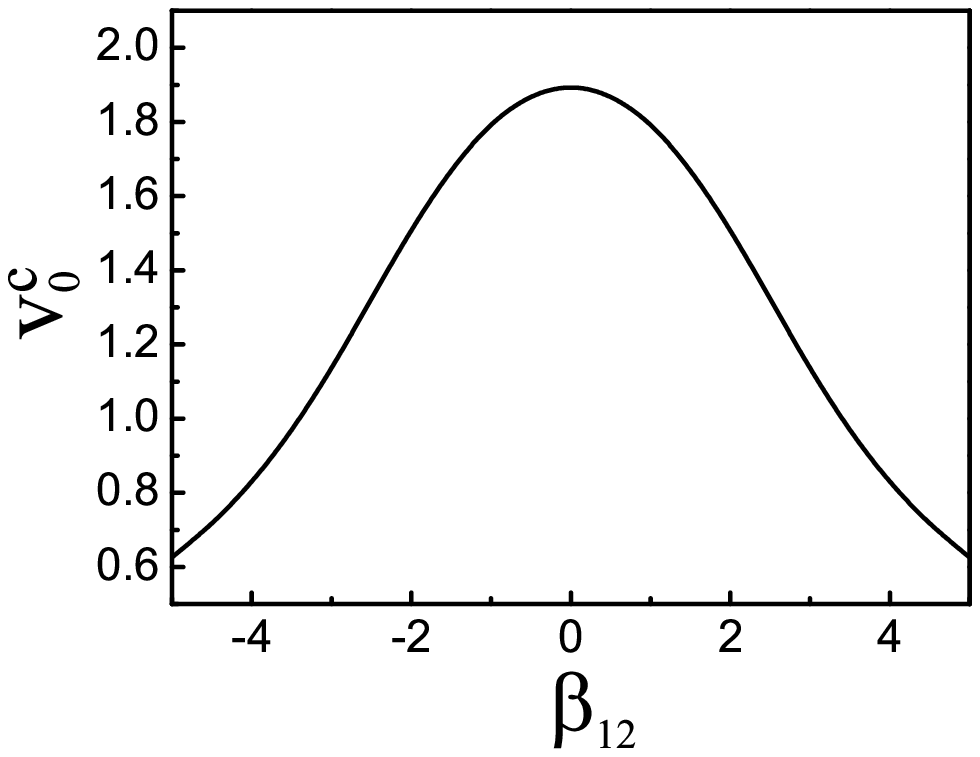}\includegraphics[scale=0.42]{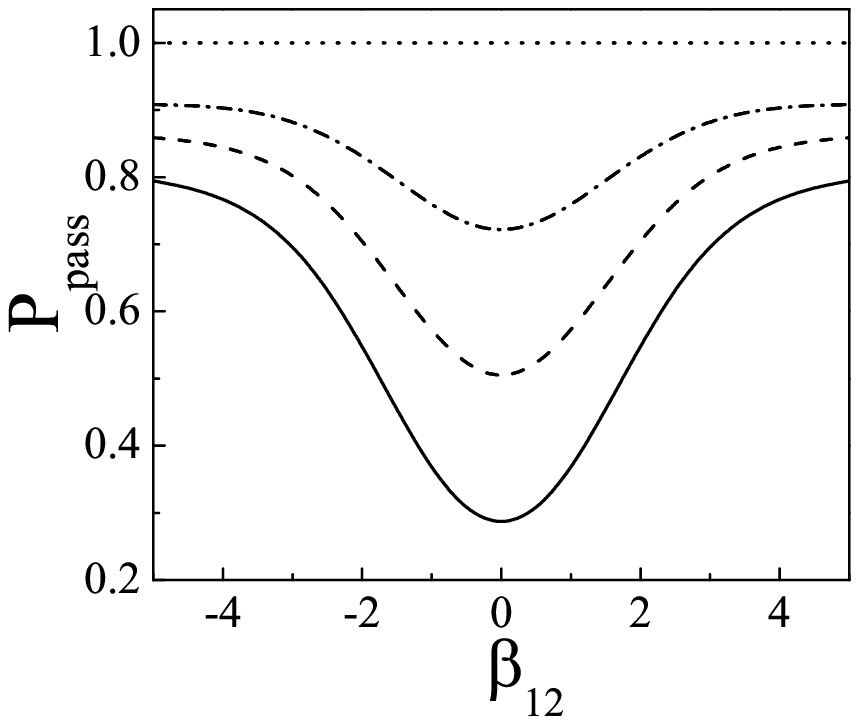}
\includegraphics[scale=0.4]{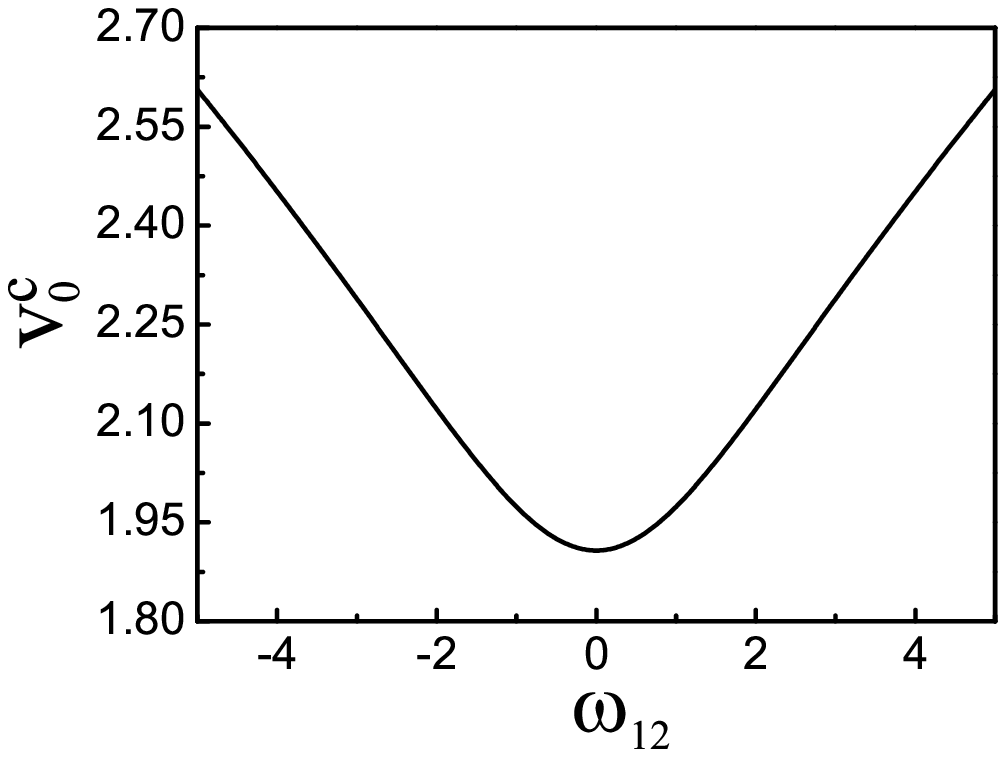}\includegraphics[scale=0.41]{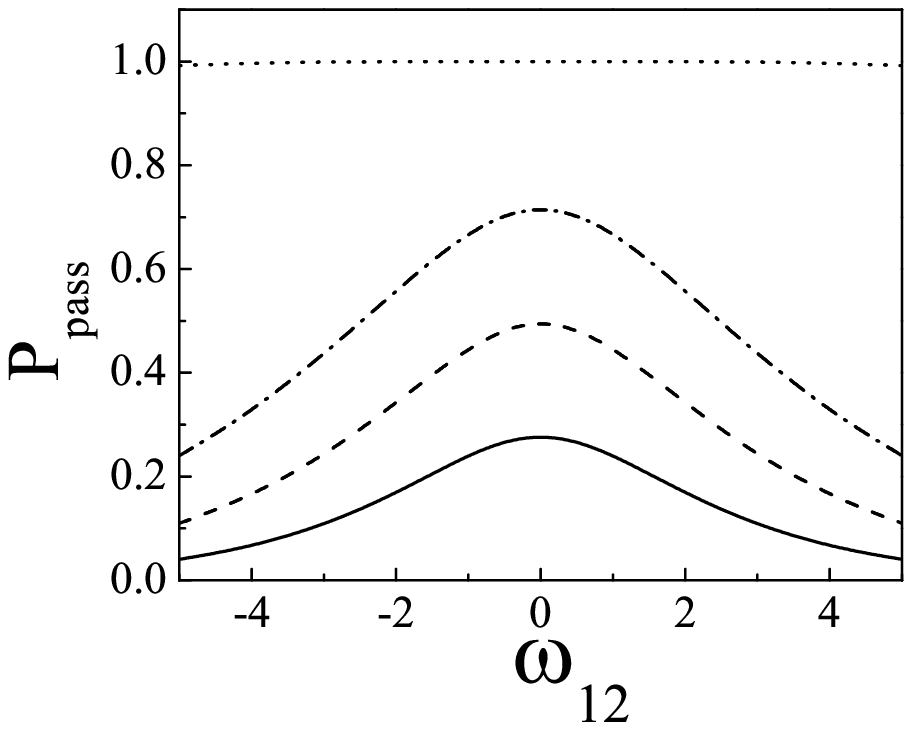}
\caption{The critical velocity (left) and the stationary passing
probability (right) as functions of various off-diagonal components
$m_{12}$, $\beta_{12}$, and $\omega_{12}$, respectively. The
parameters used are: $m_{11}=1.5$, $m_{22}=2.0$, $\beta_{11}=1.8$,
$\beta_{22}=1.2$, $\omega_{11}=-2.0$, $\omega_{22}=1.5$, and
$\theta=0$. The initial velocities of the particle  are
$v_{0}=4.0,2.2,1.9,1.6$ from top to bottom (right).}
\end{figure}

As one has known in the 1D case, the passing probability increases
from $0$ to $1$ when the initial velocity of the particle increases.
The critical velocity is defined by the passing probability being
equal to $\frac{1}{2}$. This leads to the condition:
$\lim_{t\rightarrow\infty}\langle x_{1}(t)\rangle=0$. If all the
off-diagonal components of the three coefficient tensors are not
considered, as well as $x_{20}$ and $v_{20}$ are taken to be zero,
the critical velocity is determined from Eq. (6):
$v^{c}_{0}=\left[\textsl{F}_{1}(a)/\textsl{F}_{3}(a)\right]x_{10}$,
where $a$ is the largest positive root of $\textsl{P}(s)=0$. This is
in fact identical to the one-dimensional result:
$v^c_{10}=-x_{10}(\sqrt{\beta^2_{11}+4\omega_{11}}+\beta_{11})/(2m_{11})$
\cite{abe1}, which is proportional to the friction strength.

We now consider all the off-diagonal components of three coefficient
tensors, namely, the correlations of two degrees of freedom are
taken into account, the critical velocity can also be determined by
$\lim_{t\rightarrow\infty}\langle x_{1}(t)\rangle=0$ and results in
\begin{eqnarray}
v^{c}_{_{0}}=-\frac{C_{1}(\infty)x_{10}+C_{2}(\infty)x_{20}}
{C_{3}(\infty)\cos\theta+C_{4}(\infty)\sin\theta},
\end{eqnarray}
where  $\theta$ denotes the incident angle between the initial
velocity and the $x_1$-direction, hence
$v_{10}=v_{0}\textrm{cos}\theta$ and
$v_{20}=v_{0}\textrm{sin}\theta$.

Quantities plotted in the figures 1-7 are dimensionless and
$k_{_{B}}=1.0$ except for the units having been included in the
figure caption. In Fig. 1, we plot the critical velocity and the
stationary passing probability as functions of the off-diagonal
components of three coefficient tensors, where one of off-diagonal
components varies and the other two  are fixed to be zero. The
stationary passing probability is calculated by $
P_{\textmd{pass}}=\lim_{t\to\infty}\frac{1}{2}\textmd{erfc}\left[-\langle
x_{1}(t)\rangle/(\sqrt{2}\sigma_{x_1}(t))\right]$. It is shown that
the critical velocity increases with increasing the absolute value
of $m_{12}$ or $\omega_{12}$; while decreases with the increase of
$|\beta_{12}|$. The larger critical velocity of a system needs, the
more difficult for the particle is to arrive at the potential top.
This also implies that the passing probability is small when the
dissipation along the diffusive path is large if the potential
differences between the saddle point and the initial positions are
equivalent. As is shown in the figure, the behavior of the passing
probability is opposite to that of the corresponding critical
velocity.

\begin{figure}
\includegraphics[scale=0.8]{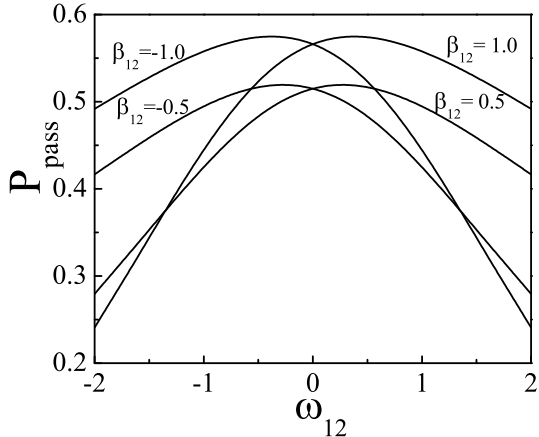}
\caption{The stationary passing probability as a function of the
off-diagonal component $\omega_{12}$ for various $\beta_{12}$. The
parameters used are the same as those in the figure 1.}
\end{figure}

Figure 2 shows the stationary passing probability in the presence of
two off-diagonal components $\omega_{12}$ and $\beta_{12}$,
simultaneously, for $m_{12}=0$.  It is seen that the maximum of the
passing probability does not appear in the vertical case
($\omega_{12}=0$). In the 2D PES, the particle is usually supposed
to travel along the potential valley and then the steepest decedent
direction. Because this is the direction which faces a smaller
potential barrier. However, it may not be a path with a weaker
damping. Under the effect of the off-diagonal component of the
friction tensor, the particle is forced to select a better path with
both low potential barrier and weak friction to surmount the saddle
point of the potential.

\subsection{Determination of the optimal path}

\begin{figure}
\includegraphics[scale=0.8]{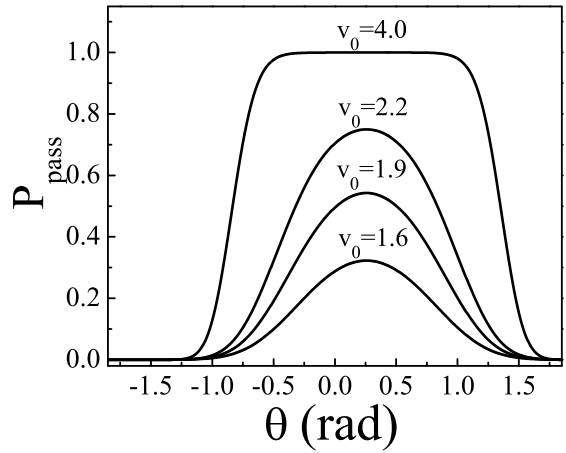}
\caption{The stationary passing probability as a function of the
incident angle. Here $m_{12}=0.6$, $\beta_{12}=0.8$, and
$\omega_{12}=-0.5$, as well as the other parameters are used as
those in the figure 1.}
\end{figure}

Where is the optimal incident direction which enables the particle
with given initial kinetic energy to have a larger passing
probability? In order to determine this direction, we need to choose
a special angle $\theta_{m}$ which enables the critical velocity to
reach its minimum, i.e., from Eq. (7),
\begin{equation}
 \frac{dv^c_0}{d\theta}|_{\theta=\theta_m}=0,\hspace{0.5cm}\theta_{m}=\textrm{arctan}
 \left(\frac{C_{4}(\infty)}{C_{3}(\infty)}\right).
\end{equation}
 In fact, the largest analytically root  of $\textsl{P}(s)=0$ dominantly
determines the passing probability. The optimal incident angle
$\theta_{m}$ can then be expressed by the Langevin coefficients as
\begin{eqnarray}
\theta_{m}=\textrm{arctan}\left(\frac{m_{12}(\beta_{22}a+\omega_{22})-m_{22}(\beta_{12}a+\omega_{12})
}{m_{11}\textsl{F}_{5}(a)+m_{12}\textsl{F}_{6}(a)}\right).
\end{eqnarray}
Using the same parameters as those written in Figs. 1 and 3, we
obtain $\theta_{m}\simeq 0.258$ rad, as is explicitly shown in Fig.
3, corresponding to the maximum of the stationary passing
probability.

\begin{figure}
\includegraphics[scale=0.6]{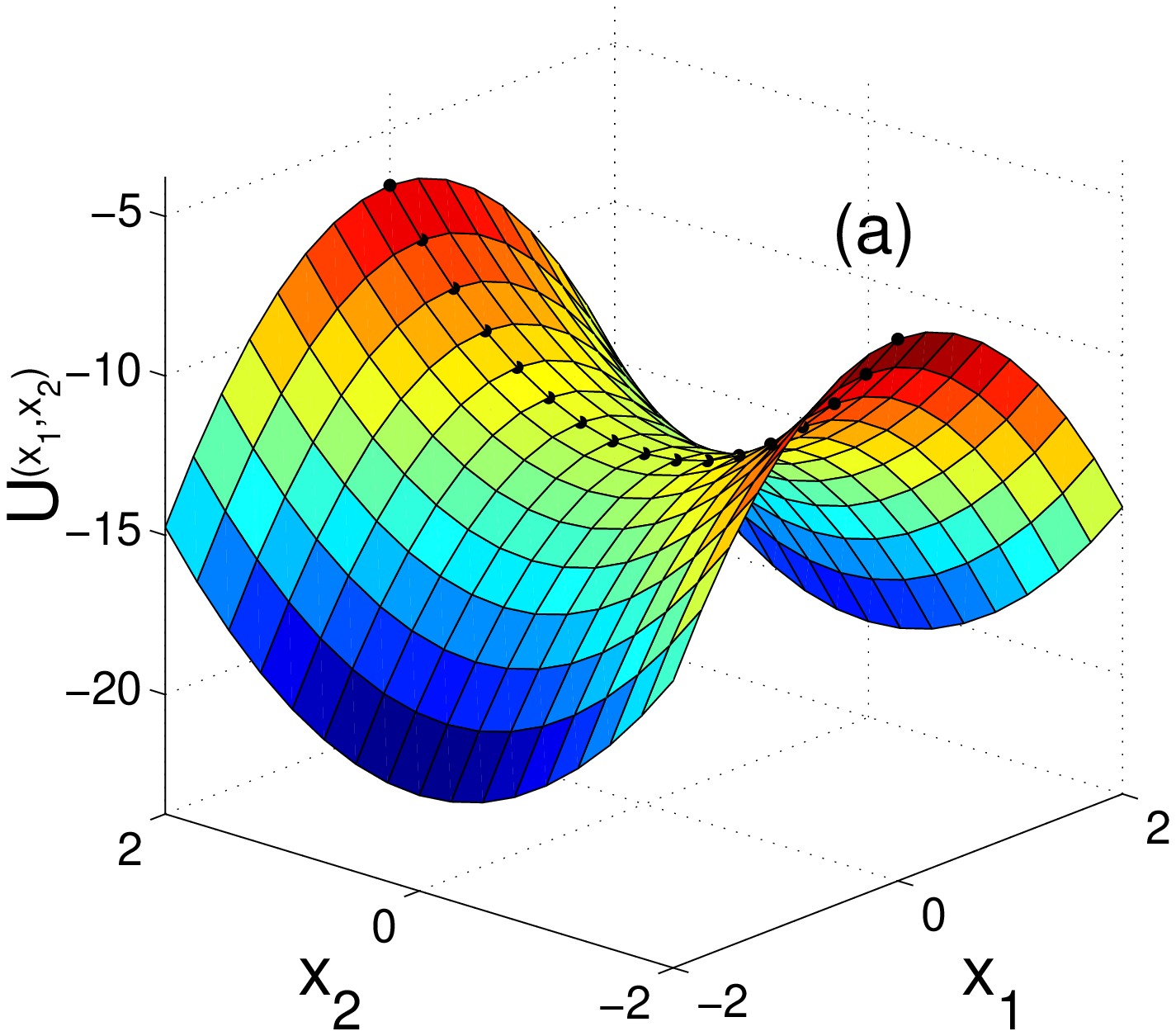}
\includegraphics[scale=1.]{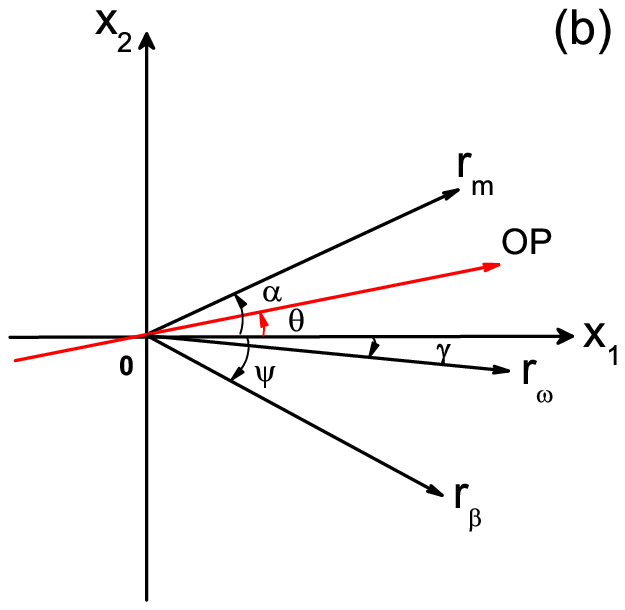}
\caption{(color online.) (a) The 2D potential energy surface, where
the dotted curve  is the saddle ridge line. (b) A schematic
illustration of  the optimal diffusion path (OP), where
$\bf{r}_{m}$, $\bf{r}_{\beta}$, and $\bf{r}_{\omega}$ denote the
major axes of the inertia, friction, and potential-curvature
tensors, respectively. }
\end{figure}

We now define
 $\gamma$, $\psi$ and $\alpha$ to be the rotation angle of the major
axis of the potential-curvature, friction and inertia tensors,
respectively. They are found to have the following expressions:
\begin{eqnarray}
\textrm{tan}2\gamma&=&\frac{2\omega_{12}}{\omega_{22}-\omega_{11}},
\hspace{0.5cm}\textrm{tan}2\psi=\frac{2\beta_{12}}{\beta_{22}-\beta_{11}},\nonumber\\
 \textrm{tan}2\alpha&=&\frac{2m_{12}}{m_{22}-m_{11}}.
\end{eqnarray}
  As an example, for the case we have studied in Figs. 1 and 3, these
angles are: $\gamma\simeq -7.973^{\circ}$, $\psi\simeq
-34.722^{\circ}$, and $\alpha\simeq 33.690^{\circ}$. For comparison,
we write the optimal incident direction of the the particle we have
obtained in the unit of one degree, i.e., $\theta_{m}\simeq
14.779^{\circ}$ (0.258 rad).

In Fig. 4, we plot the two-dimensional quadratic potential and the
optimal path in the $x_1$-$x_2$ plane in a way of schematic
illustration. All the coefficient elements used here have been
written in the figures 1 and 3. It is illustrated that the direction
of the optimal path
 departs from the $x_1$-direction. The effect of off-diagonal
component of the inertia tensor makes the average path of the
diffusive system turn toward the positive $x_2$-axis, while the
off-diagonal component of friction leads the mean path of the
particle toward the negative $x_2$-axis. Finally, the competition of
these two effects results in the optimal diffusive path shown in
Fig. 4 (b). This phenomenon is similar to the quasi-stationary flow
passing over the barrier in the fission case \cite{zhang}, where the
magnitude of the current is strongly influenced  by the off-diagonal
components of inertia and friction tensors.

\begin{figure}
\includegraphics[scale=0.8]{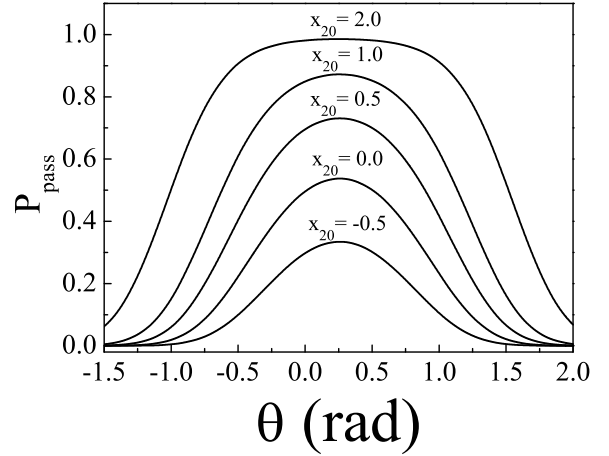}
\caption{The stationary passing probability as a function of
$\theta$ for various  $x_{20}$. Here $x_{10}=-1.0$, $v_{0}=1.9$,
$m_{12}=0.6$, $\beta_{12}=0.8$, and $\omega_{12}=-0.5$, as well as
the other parameters are the same as the figure 1.}
\end{figure}

Figure 5 shows the dependence of the stationary passing probability
on the incident angle of the particle starting from various initial
positions but with fixed initial kinetic energy. It is seen that the
passing probability of the particle starting from a large positive
$x_{20}$ position is larger than that of starting from both small
 and negative $x_{20}$ positions.This is because the energy
difference between the potential top and the initial position of the
particle is small for the former. Amusingly, we find that the
difference between the passing probabilities of two symmetrical
positions $(-1.0,-0.5)$ and $(-1.0,0.5)$ is observably large.

\begin{figure}
\includegraphics[scale=0.4]{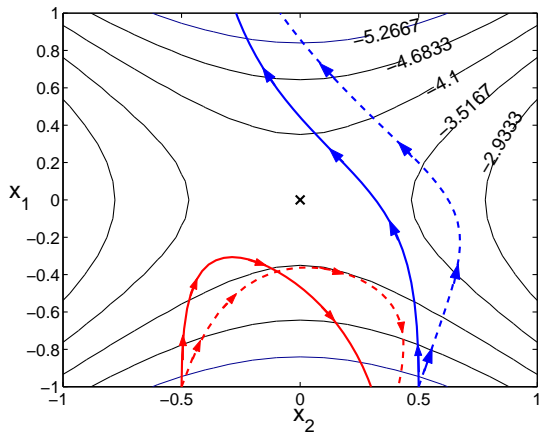}
\caption{(color online.)  The  mean diffusion path of a particle
starting from the initial positions of two kinds  at fixed
$v_0=1.9$, where the solid and dashed lines correspond to $\theta
=0$ and $\theta=0.258$ rad, respectively. Here all the Langevin
parameters are the same as the figures 1 and 5.}
\end{figure}

\begin{figure}
\includegraphics[scale=0.7]{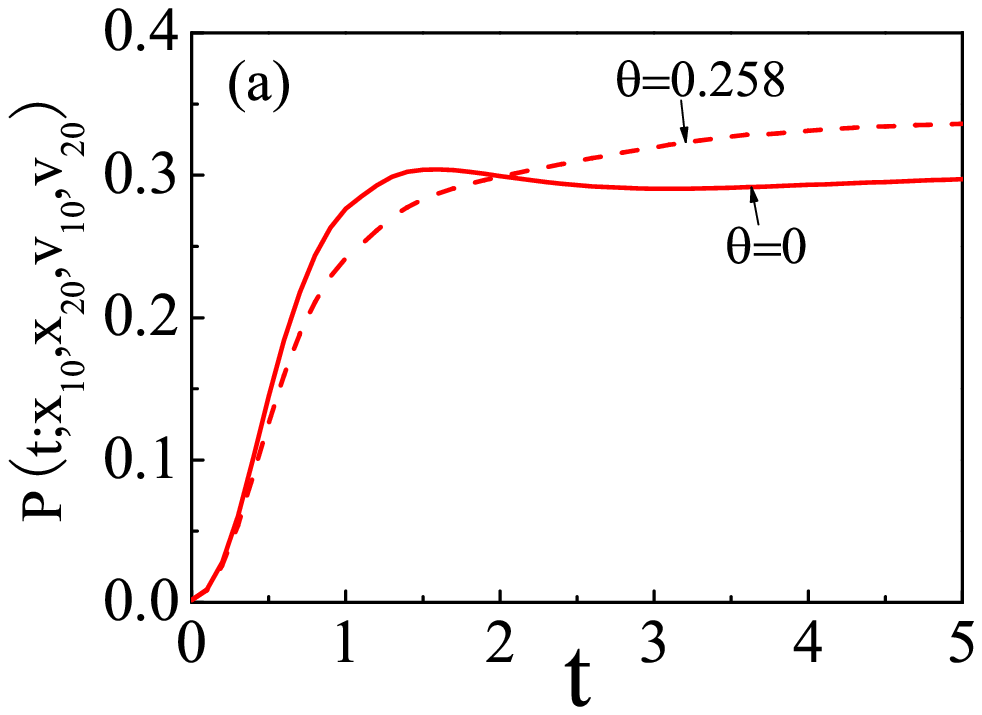}
\includegraphics[scale=0.7]{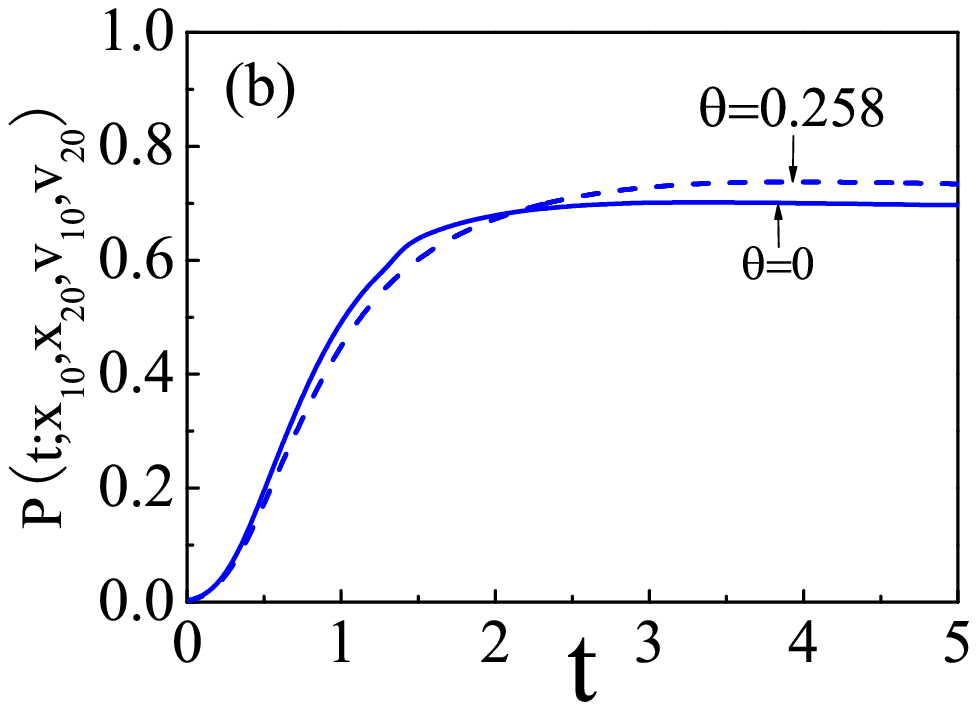}
\caption{(color online.) Time-dependent passing probability  for
various $x_{20}$ and $\theta$. Here $x_{10}=-1.0$, $v_{0}=1.9$,
$x_{20}=-0.5$ in (a) and $x_{20}=0.5$ in (b), as well as the
Langevin parameters are the same as the figures 1 and 5.}
\end{figure}

For a clearly understanding of the above results, we plot in Fig. 6
the mean diffusive path of a particle starting from different
initial positions with different incident angles. Here $\langle
x_1(t)\rangle$ has been obtained in Eq. (6) and
\begin{eqnarray}
\langle
x_{2}(t)\rangle=\sum^{2}_{i=1}\left[C_{i+4}(t)x_{i0}+C_{i+6}(t)v_{i0}\right],
\end{eqnarray}
where all the time-dependent quantities are given in the appendix.
The critical velocities are calculated by using Eq. (7):
$v_0^c=1.5791$ when $x_{20}=0.5$; $v_0^c=2.2321$ when $x_{20}=-0.5$,
for $x_{10}=-1.0$ and $\theta=0$. Hence the stationary passing
probability of the particle starting from $x_{20}=0.5$ is larger
than that of the particle starting from $x_{20}=-0.5$. In
particular, under the present circumstance, the diffusive process of
the particle with different incident angles shows an interesting
behavior. Because the initial velocity of the particle along the
$x_1$-direction for $\theta=0$ is larger than that of $\theta=0.258$
rad, the former can move to a position being closer to the saddle
point than the latter. Thus the passing probability for $\theta=0$
is larger than that of $\theta=0.258$ rad at the beginning. However,
the width of the Gaussian distribution is independent of the
incident angle and increases with the increase of time. Although the
center position of the particle's distribution with $\theta=0.258$
rad is behind that of $\theta=0$, as the time goes on, it will have
a lager share of distribution passed the saddle point. Therefore,
the passing probability for a particle with incident angle
$\theta=0.258$ rad is larger than that of $\theta=0$ in the long
time.

Time-dependent passing probabilities shown in Figs. 7 (a) and 7 (b)
are also in complete agreement with the above theoretical analysis.

\section{Application to fusion of massive nuclei}

We now apply the present 2D simplified diffusive model to
investigate the fusion of two massive nuclei, which has been
described by directional diffusion  over the saddle point
\cite{abe1}. As a particular example, we calculate the fusion
probability of the nearly symmetrical reaction system
$^{100}\textrm{Mo}+^{110}\textrm{Pd}$ \cite{sch}, which is plotted
as a function of the center-of-mass energy $E_{c.m.}$ in Fig. 8. A
schematic illustration of the deformation of the compound nucleus is
also shown in this figure. The temperature of the fusing system is
determined by $aT^{2}=E_{c.m.}+Q-E_{B}$, where $a=A/10$ is the
energy level constant with $A$ the nucleon number of the compound
nucleus, $Q$ denotes the reaction  $Q$ value and $E_{B}$ the barrier
height of fission potential.

The $\{c,h,\alpha\}$ shape parametrization \cite{brac} with the
elongation $c$ (the half of the nuclear length) and neck variable
$h$ are used, i,e, $x_1=c$, $x_2=h$, and the asymmetrical parameter
$\alpha$ is fixed to be zero. The inertia and friction tensors are
calculated by the Werner-Wheeler method and the one-body dissipative
mechanism \cite{jia}, respectively, all the Langevin coefficients
are considered to be constants at the saddle point. The three
components of potential-curvature tensor are: $\omega_{11} =
-28.2304$, $\omega_{22} = 275.4211$, and $\omega_{12} = 50.4551$ in
the unit of MeV; the components of friction tensor $\beta_{11} =
701.9967$, $\beta_{22}= 621.4425$, $\beta_{12} = 601.4934$ in the
unit of $10^{-21}$ MeV$\cdot$sec; the inertia elements $m_{11} =
102.4081$, $m_{22} = 134.4673$, and $m_{12} = 110.3783$ in the unit
of $10^{-42}$ MeV$\cdot \textmd{sec}^{2}$.

\begin{figure}
\includegraphics[scale=0.8]{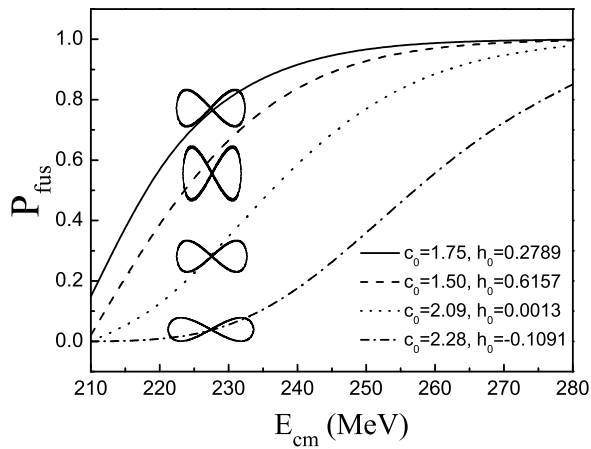}
\caption{The fusion probability of the reaction
$^{100}\textrm{Mo}+^{110}\textrm{Pd}$ as a function of the
center-of-mass energy for various initial positions. Schematically
illustrated as well are the deformed shape of the compound nucleus.}
\end{figure}

It is found a highlighted interesting result from Fig. 8 that there
exists the optimal collision shape for projectile and target nuclei,
which induces the maximum fusion probability under the same
center-of-mass energy. It can be easily understood from a combining
role of the off-diagonal components of three dynamical coefficient
tensors. This concludes that the fusion probability of massive
nuclei can be enhanced if the two collision heavy ions are polarized
to be ellipsoid and the collision direction between the long and
short axis of ellipsoid is appropriately selected.  For the fusion
of deformed massive nuclei, there exists the optimal angle for the
incident nucleus to collide with the target one, which favors the
fusion of heavy ions to be accomplished.

Figure 9 shows the fusion probability of
$^{100}\textrm{Mo}+^{110}\textrm{Pd}$ as a function of the
center-of-mass energy when the off-diagonal components are
considered partly. Here the initial position of the fusing system is
chosen into the optimal one, i.e., $c_0=1.75$ and $h_0=0.2789$ as
well as all the parameters used are the same as the figure 8. This
fusion reaction system is regarded as an example to compare the
results with and without off-diagonal components in the potential
surface, friction and mass parameters, which is reflected in the
result presented in the above section. In fact, the case of without
all off-diagonal components is equivalent to the one dimension  or
without the neck situation \cite{agu1}.

It is seen from Fig. 9 that the increase of the 1D fusion
probability curve  versus the energy is faster than that of 2D case.
It has been known that the pervious 1D diffusion model without the
neck variable proposed the fusion probability larger than the
experimental data, so the present completely coupled 2D diffusion
model might be appropriate. Moreover, the results for the presence
of only one of three off-diagonal components can also be understood
from the critical velocity (kinetic energy), to see Fig. 1. Namely,
the larger the critical kinetic energy of the system is, the less
the fusion probability is for the same center-of-mass energy. The
nonvanishing $\beta_{12}$ allows the smallest critical kinetic
energy and the presence of $m_{12}$ leads to the largest critical
kinetic energy. Therefore,  we have a relation for the fusion
probabilities: $P_{\textmd{fus}}(\beta_{12}\neq
0)>P_{\textmd{fus}}(\omega_{12}\neq 0)>P_{\textmd{fus}}(m_{12}\neq
0)$ at a fixed center-of-mass energy.

\begin{figure}
\includegraphics[scale=0.8]{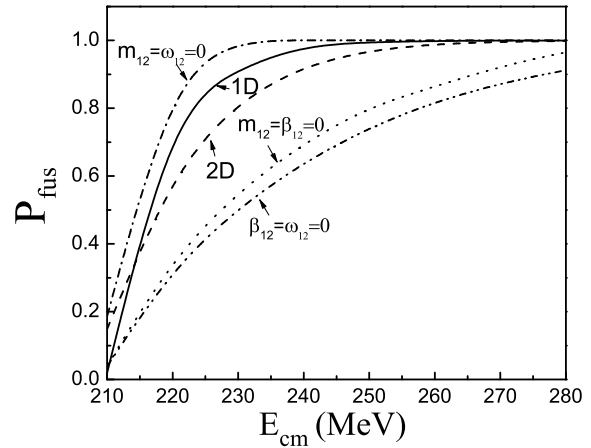}
\caption{The fusion probability of the reaction
$^{100}\textrm{Mo}+^{110}\textrm{Pd}$ as a function of the
center-of-mass energy for the situations with and without
off-diagonal components.}
\end{figure}

\section{SUMMARY}

We have studied the diffusion process of a particle passing over the
saddle point of a two-dimensional non-orthogonal quadratic
potential. The expression of the passing probability is obtained
analytically, where the inertia and friction tensors are not
diagonal. The optimal incident angle of the particle's initial
velocity is determined. Our results have shown that the optimal
diffusive path, which departs from the potential valley in the
two-dimensional potential energy surface, induces the maximum
saddle-point passing probability. This is due to the competition
effect between the off-diagonal components of inertia, friction and
potential-curvature tensors. We have investigated the fusion
probability of massive nuclei and compared the results with and
without off-diagonal terms, for instance, the reaction of
$^{100}\textrm{Mo}+^{110}\textrm{Pd}$. Due to the influences of
off-diagonal components of inertia and friction upon the diffusive
path, which are calculated by the $\{c, h, \alpha\}$
parametrization, the fusion probability  can be enhanced for an
appropriate choice of the collision direction of the deformable
target and projectile nuclei. The optimal configurations of
colliding nuclei is between spherical and extremely deformed ones.
In further, the present study also provides useful information in
connection with the synthesis of superheavy elements.

\section * {ACKNOWLEDGEMENTS}
 This work was supported by the National
Natural Science Foundation of China under Grant Nos. 10674016,
10747166  and the Specialized Research Foundation for the Doctoral
Program of Higher Education under Grant No. 20050027001.

\section * {APPENDIX. THE EXPRESSIONS OF $\langle x_1(t)\rangle$ AND $\langle x_2(t)\rangle$}

The quantities appear in the expression of $\langle x_1(t)\rangle$
are
\begin{eqnarray}
\textsl{P}(s)&=&(\textrm{det}m)s^{4}+(m_{11}\beta_{22}+m_{22}\beta_{11}-2m_{12}\beta_{12})s^{3}
\nonumber\\&&
+(\textrm{det}\beta+m_{11}\omega_{22}+m_{22}\omega_{11}-2m_{12}\omega_{12})s^{2}\nonumber\\&&
+(\beta_{11}\omega_{22}+\beta_{22}\omega_{11}-2\beta_{12}\omega_{12})s+\textrm{det}\omega,\nonumber\\
\textsl{F}_{1}(s)&=&(\textrm{det}m)s^{3}+(m_{11}\beta_{22}+m_{22}\beta_{11}-2m_{12}\beta_{12})s^{2}
\nonumber\\&&
+(\textrm{det}\beta+m_{11}\omega_{22}-m_{12}\omega_{12})s
+\beta_{11}\omega_{22}-\beta_{12}\omega_{12},\nonumber\\
\textsl{F}_{2}(s)&=&(m_{12}\omega_{22}-m_{22}\omega_{12})s+\beta_{12}\omega_{22}-\beta_{22}\omega_{12},\nonumber\\
\textsl{F}_{3}(s)&=&(\textrm{det}m)s^{2}+(m_{11}\beta_{22}-m_{12}\beta_{12})s
+m_{11}\omega_{22}\nonumber\\
&&-m_{12}\omega_{12},\nonumber\\
\textsl{F}_{4}(s)&=&(m_{12}\beta_{22}-m_{22}\beta_{12})s
+m_{12}\omega_{22}-m_{22}\omega_{12},\nonumber\\
\textsl{F}_{5}(s)&=&m_{22}s^{2} +\beta_{22}s+\omega_{22},\nonumber\\
\textsl{F}_{6}(s)&=&-m_{12}s^{2}-\beta_{12}s-\omega_{12},
\end{eqnarray}
where $\det m=m_{11}m_{12}-m^2_{12}$ and
$\det\beta=\beta_{11}\beta_{22}-\beta^2_{12}$.

The time-dependent factors in the expression of $\langle
x_2(t)\rangle$ in Eq. (11) read
 $C_{j}(t)=\textsl{L}^{-1}[\textsl{F}_{j+2}(s)/\textsl{P}(s)]$
$(j=5,\cdots,8)$ being resulted from the inverse Laplace transforms,
as well as
\begin{eqnarray}
\textsl{F}_{7}(s)&=&(m_{12}\omega_{11}-m_{11}\omega_{12})s+\beta_{12}\omega_{11}-\beta_{11}\omega_{12},\nonumber\\
\textsl{F}_{8}(s)&=&(\textrm{det}m)s^{3}+(m_{11}\beta_{22}+m_{22}\beta_{11}-2m_{12}\beta_{12})s^{2}\nonumber\\&&
+(m_{22}\omega_{11}-m_{12}\omega_{12}+\textrm{det}\beta)s+\beta_{22}\omega_{11}\nonumber\\
&&-\beta_{12}\omega_{12},\nonumber\\
\textsl{F}_{9}(s)&=&(m_{12}\beta_{11}-m_{11}\beta_{12})s+m_{12}\omega_{11}-m_{11}\omega_{12},\nonumber\\
\textsl{F}_{10}(s)&=&(\textrm{det}m)s^{2}+(m_{22}\beta_{11}-m_{12}\beta_{12})s+m_{22}\omega_{11}\nonumber\\
&&-m_{12}\omega_{12}.
\end{eqnarray}

\end{document}